\documentclass[a4paper]{article}
\usepackage{graphicx}
\newcommand{\be}{\begin{equation}}
\newcommand{\ee}{\end{equation}}
\newcommand{\bea}{\begin{eqnarray}}
\newcommand{\eea}{\end{eqnarray}}
\begin{document}
\begin{titlepage}
\title{Complementarity of Entanglement and Interference}
\author{ A.  Hosoya, A.  Carlini and S. Okano\cr Department of Physics \cr
Tokyo Institute of Technology\cr Oh-Okayama, Meguro-ku, Tokyo 152\cr
Japan} 
\maketitle
%%%%%%%%%%%%%%%%%%%%%%%%%%%%%%%%%%%%%%%%%%%%%%%%%%%%%%%%
\begin{abstract}
A complementarity relation is shown between the visibility of interference
and bipartite entanglement  in a two qubit  interferometric system when
the parameters of the quantum operation change for a given input state.  The
entanglement measure is a decreasing function of the visibility of
interference.  The implications for quantum computation are briefly
discussed.
\end{abstract}
PACS Nos: 03.65.Vf,03.65.Yz,03.67.Lx,03.67.Mn
\end{titlepage}
%%%%%%%%%%%%%%%%%%%%%%%%%%%%%%%%%%%%%%%%%%%%%%%%%%
%%%%%%%%%%%%%%%%%%%%%%%%%%%
\section{Introduction}
%%%%%%%%%%%%%%%%%%%%%%%%%%%
%introduction
The well-known complementarity or duality of particle and wave is one of
the deepest concepts in quantum mechanics. In this work we 
investigate a similar complementarity between entanglement and interference,
which are both important concepts in quantum mechanics \cite{PERE} and also
powerful tools in quantum information processing \cite{CN}.

More specifically, we  quantify the complementarity of entanglement
and interference visibility for a bipartite system within an
interferometric model.
The entanglement measures for bipartite mixed states were introduced and their
significance in quantum processing was discussed by Bennett et al.
\cite{BENNETT}.
An extensive discussion of entanglement measures can be found, e.g., in Horodecki \cite{HO}
and in Vedral and Plenio \cite{VEDRAL}.
We use the interferometric model of Sj{\"o}qvist et al.  \cite{SJ}, which introduces internal degrees of freedom for the impinging particle.
The authors of Ref. \cite{SJ} were interested
in an experimental approach (recently verified by NMR techniques \cite{JD})
to the geometric phase for mixed states.
In this model the geometric phase can be
observed by a shift of the interference pattern so that the
visibility of the latter and that of the geometric phase are the same thing.

Here we show that, for any bipartite system of path and internal degrees of freedom, 
the entanglement measure of the compound output state (the negativity in 
the two qubits mixed case and the von Neumann entropy of the reduced density operator
in the case of pure states with arbitrary dimensions) is a
decreasing function of the visibility of the interference in the counting rate of outgoing particles,
where the control parameters of the quantum operation change for a given input.

We should stress that in the present literature a number of works 
appeared dealing with topics close to ours to various extents.
Complementarity relations between single and two particle fringe visibilities were
found by Jaeger et al. \cite{JAEGER}, between distinguishability and visibility 
by Englert and collaborators \cite{ENGLERT1} etc..  
For example, in Englert et al. \cite{ENGLERT2} the
wave particle duality is discussed by showing the complementarity between the
distinguishability of "which-way" path and the visibility of
interference in Young's double slit experiment.  However, the
distinguishability seems qualitatively related to the entanglement but
not in a quantitative level.
Furthermore, e.g. Jacob and Bergou \cite{JAKOB} investigated the relation between the 
concurrence of entanglement
and visibility in a bipartite system. For pure states the authors observed
the same complementarity as the one described here in Sec. 4 , while for
mixed states they found a weaker statement in the form of an inequality
(the relationship between indistinguishability and average
predictability and coherence for an arbitrary state of two qubits was also 
recently analyzed in \cite{TESSIER}). 
Ekert et al. \cite{EKERT} presented a quantum network based on the controlled-SWAP gate, and whose interferometric setup is similar to ours, that can extract certain properties of quantum states without recourse to quantum tomography and can be used as a basic building block for direct quantum estimations of both linear and non-linear functionals of any density operator. 
Recently, Carteret \cite{CARTERET} also constructed an interesting quantum circuit
corresponding to the interferometric model which measures the trace of powers
of the partially transposed density operator, by which we can compute its
spectrum and then the concurrence.
Hartley and Vedral \cite{HARTLEY} finally
used a similar method to measure and relate the von Neumann entropy and the geometric
phase visibility for two and three dimensional Hilbert spaces.

The paper is organized as follows.
We first describe the interferometry model for a two qubit state and give a more specific definition 
for the entanglement and the visibility of interference in 
Sec. 2.  In Sec. 3 we demonstrate the complementarity of
entanglement and interference for a mixed state which has an entanglement 
boundary in the parameter space.
In Sec. 4 we consider a pure bipartite system, while Sec. 5 is devoted to the analytical 
description of the complementarity relation for a pure bipartite system in arbitrary dimensions.
Sec.  6  is finally devoted to a brief summary and a discussion
of some of the possible implications of our results on quantum computation.
%%%%%%%%%%%%%%%%%%%%%%%%%%%
\section{An Interferometry  Model}
%%%%%%%%%%%%%%%%%%%%%%%%%%%%
We consider the model of Sj\"oqvist et al. \cite{SJ}, which is essentially a standard
Mach-Zehnder interferometer where the Hilbert space consists of the path states for 
the impinging particle (e.g., a neutron) and of the states for its internal degrees of freedom
(e.g., the spin) (see Figs. 1-2). A
detector is set at the output to detect particles moving in the
horizontal direction.  By means of a  phase shifter located along one of the
horizontal paths, we can see an interference phenomenon in the
counting rate of the horizontally outgoing particles as a function of the phase shift.
We are then going to study the relation between the visibility of this interference pattern
and a measure of the entanglement between path and internal output states.
%%%%%%%%%%%%%%%%%%%%%%%%%%%%%%%%%%%%%
\begin{figure}
\includegraphics[width=.9\linewidth]{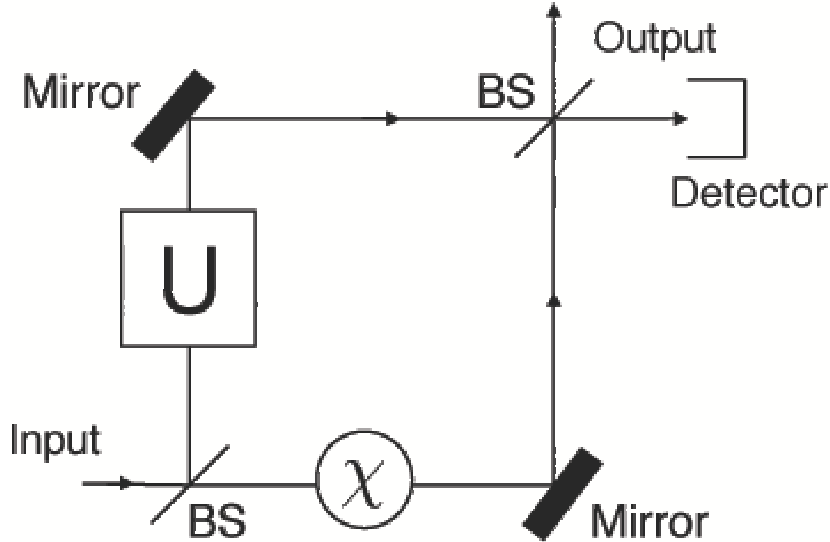}
\caption{Sj\"oqvist's interferometer model.
Particles (spin: 1/2) entering  the interferometer along the horizontal 
direction are globally phase-shifted 
by $\chi $ and their spin states are transformed by the unitary $U$.
Then the particles are reflected by two mirrors and merge through a beam 
splitter, after which they are detected in the horizontal direction.}
\end{figure}

\begin{figure}
\includegraphics[width=1.1\linewidth]{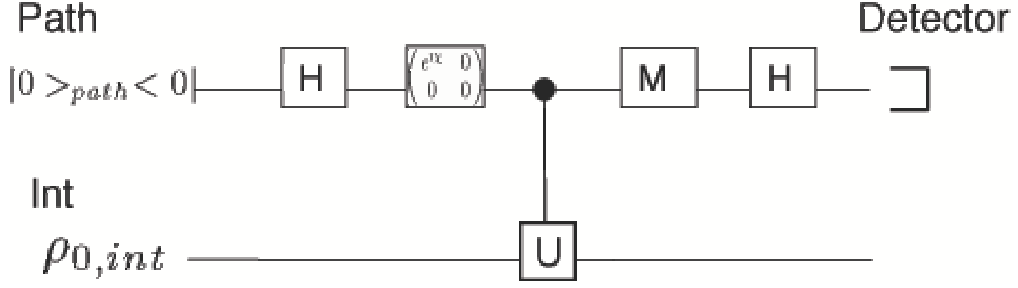}
\caption{Quantum circuit equivalent to Sj\"oqvist's interferometer 
model. }
\end{figure}
%%%%%%%%%%%%%%%%%%%%%%%%%%%%%%%%%%%%%%

More in detail, the state of the compound system ``path-internal states'' will
change as 
\be 
\rho_{IN}\rightarrow \rho_{OUT}=V\rho_{IN}V^{\dagger},
\ee 
where the unitary operator $V$ is given by 
\be
V\equiv V_{H}V_{M}V_{C}V_{H}, 
\ee 
$V_{H}$ represents the unitary
transformation corresponding to the beam splitters,
\bea
V_{H}&\equiv &H\otimes I,\nonumber\\
H&\equiv &{1\over \sqrt{2}}\left(\begin{array}{clcr}1 & ~~1 \\ 1 &
-1\end{array}\right), 
\eea 
while $V_{M}$ is the unitary transformation
corresponding to the mirrors,
\bea 
V_{M}&\equiv &M\otimes I,\nonumber\\
M&\equiv &\left(\begin{array}{clcr}0 & 1 \\ 1 & 0\end{array}\right).
\eea 
Note that neither of these operators affect the internal states.  The
operator $V_{C}$, instead, also acts on the internal space depending
on the path travelled by the particle.  If the particle passes through
the vertical path its internal state undergoes a unitary
transformation $U$, while if it passes through the horizontal path it
picks up a phase $\chi$.  Explicitly, 
\be
V_{C}\equiv \left(\begin{array}{clcr}0 & 0\\ 0 & 1\end{array}\right)\otimes
U+\left(\begin{array}{clcr}e^{i\chi}& 0 \\0&
0\end{array}\right)\otimes I.  
\ee 
The input state is chosen as a
separable state, 
\be 
\rho_{IN}=|0>_{path}<0|\otimes \rho_{0, int}, 
\ee
describing a particle entering the interferometer device along the
horizontal direction ($|0>_{path}$) and with initial internal state
$\rho_{0, int}$.  Later we take this as represented by a one qubit (spin 1/2)
system and consider two typical cases: a pure state $|0>_{int}<0|$ and the
general mixed state with the standard Bloch parametrization, 
\be
\rho_{0, int}\equiv {1\over 2}(I+\vec{b}_0\cdot\vec{\sigma}).
\label{bloch}
\ee
The output state is in general entangled and can be
expanded as 
\bea 
\rho_{OUT}&=&V\rho_{IN}V^{\dagger}\nonumber\\
&=&{1\over 4}\biggl[\left(\begin{array}{clcr} 1 & 1 \\ 1 &
1\end{array}\right)\otimes
U\rho_{0, int}U^{\dagger}+\left(\begin{array}{clcr} ~~1 & -1 \\ -1 &
~~1\end{array}\right)\otimes \rho_{0, int}\nonumber\\
&+&e^{-i\chi}\left(\begin{array}{clcr} 1 & -1 \\ 1 &
-1\end{array}\right)\otimes U\rho_{0, int}+e^{i\chi}\left(\begin{array}{clcr} ~~1
& ~~1 \\ -1 & -1\end{array}\right)\otimes \rho_{0, int}U^{\dagger}\biggr ].
\label{out1}
\eea 
The probability to detect the particle in the horizontal
direction is given by
\bea 
P_{|0>_{path}}&=&Tr_{path, int}(|0>_{path}<0|\otimes\rho_{OUT})\nonumber\\
&=&{1\over 4}Tr_{int}[ U\rho_{0, int}U^{\dagger}+
\rho_{0, int}+e^{i\chi}\rho_{0, int}U^{\dagger}+e^{-i\chi}U\rho_{0, int}]\nonumber\\
&=&{1\over 2}[1+|Tr_{int}(U\rho_{0, int})|\cdot \cos(\chi-arg(Tr_{int}(U\rho_{0, int}))]
\nonumber \\
&\equiv &{1\over 2}[1+\Gamma\cos (\chi -\gamma_g)].
\label{int1}
\eea
This exhibits an interference pattern in the $\chi$ space with the
visibility $\Gamma\equiv |Tr_{int}(U\rho_{0, int})|$ and the shift of the pattern
$\gamma_{g}\equiv arg[Tr_{int}(U\rho_{0, int})]$.  The latter one is also called the
geometric phase for a mixed state $\rho_{0, int}$ (Fig.  3).
We would like to stress here that in this interferometric model we start with a separable
state and entanglement is achieved through the selective operation $U$ acting
on the internal space. 

%%%%%%%%%%%%%%%%%%%%%%%%%%%%%%%%
%%%%%%%%%%%%%%%%%%%%%%%%%%%%%
\begin{figure}
\begin{center}
\includegraphics[width=.6\linewidth] {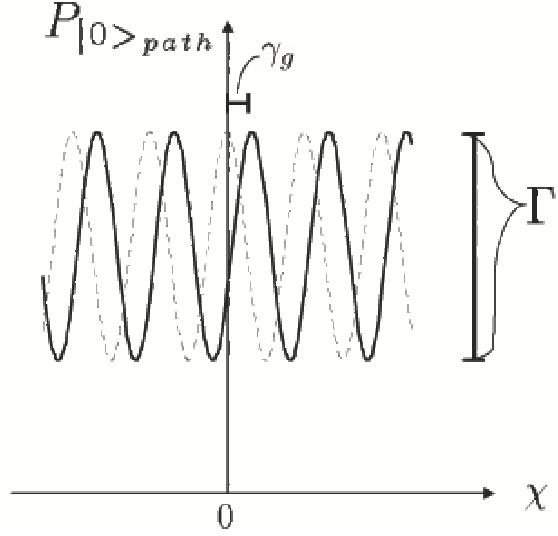}
\end{center}
\caption{
The probability to detect a particle along the horizontal direction at the 
output of the interferometer 
shows an interference pattern as a function of the phase shift $\chi $.}
\end{figure}
%%%%%%%%%%%%%%%%%%%%%%%%%%%%%%%%%%
%%%%%%%%%%%%%%%%%%%%%%%%%%%%%%%%

%%%%%%%%%%%%%%%%%%%%%%%%%%%
\section{Entanglement of  Mixed State vs. Visibility of Interference}
%%%%%%%%%%%%%%%%%%%%%%%%%%%%
%%%%%%%%%%%%%%%%%%%%%%%%%%%
\subsection{Parametrization}
%%%%%%%%%%%%%%%%%%%%%%%%%%%%
The density matrix can be explicitly written via the phase $\chi$ and
the parametrizations of the initial spin state $\rho_{0, int}$ as given
in eq.  (\ref{bloch}) and that of the unitary transformation $U$ for the
spin state as 
\bea 
&U=\left(\begin{array}{clcr} t+iz & ix+y \\ ix-y &
t-iz\end{array}\right)\;\; ;~~~ t^2+x^2+y^2+z^2=1.
\label{uparam}
\eea
As a whole we have seven parameters (${\vec b}_{0}$, $\chi$ and
$t,x,y,z$ with a constraint) in the output state $\rho_{OUT}$.  This
implies that the density operator output by the interferometer cannot
represent a two qubit mixed state in full generality.  To realize a
completely general density operator one needs nine parameters (apart from
the local degrees of $SU(2)\times SU(2)$), which can be achieved by
introducing two extra CP operations \footnote{Here CP stands for completely positive
map, a linear map $T$ between the algebras $L$ of bounded operators
in separable Hilbert spaces $N, M$, i.e. $T: L(N) \rightarrow  L(M)$, 
transforming nonnegative operators into nonnegative ones and such that
for any additional Hilbert space (of arbitrary dimension) $P$ also 
$T\otimes I_{L(P)}: L(N\otimes P)\rightarrow L(M\otimes P)$ transforms
nonnegative operators into nonnegative ones.
CP maps mathematically describe physically irreversible processes
which are typical of open systems.} along the interferometer axes (see
Sec.  3.3).

Without loss of generality the input state for the second qubit can be
taken as in eq.  (\ref{bloch}) with $\vec{b}_0=b_0\vec{e}_z$, where
the $\vec{e}_i$'s are the Bloch sphere orthonormal basis vectors.  The output
state $\rho_{OUT}$ is given by eq.  (\ref{out1}), with the last two
terms representing the interference effect, which can be seen more
explicitly by the probability to detect the particle in the horizontal
direction, according to eq.  (\ref{int1}).  In our parametrization
$Tr_{int}(U\rho_{0, int})=t+ib_{0}z$, so that the visibility $\Gamma$ is 
\be
\Gamma=\sqrt{t^2+b_{0}^2\,z^2},
\label{visibility}
\ee 
while the geometric phase $\gamma_{g}$ is
\be
\gamma_{g}=\arctan\left({b_{0}z\over t}\right ).  
\label{geomphase} 
\ee

%%%%%%%%%%%%%%%%%%%%%%%%%%
 %%%%%%%%%%%%%%%%%%
\subsection{Negativity}
%%%%%%%%%%%%%%%%%%%%%%%%%%%%
We choose the negativity as a  measure of entanglement since it is known to be LOCC
monotone \cite{VW} and it is easy to compute.
The negativity  $N(\rho) $ is defined as
\be
N(\rho)=2\sum_{\lambda_i<0}|\lambda_{i}|, 
\ee 
where the sum is over all negative
eigenvalues $\lambda_{i}$'s of $\rho^{PT}$, the partial transposition
of the density operator $\rho$.  If any of the $\lambda_{i}$'s is
negative, the state is entangled according to the criterion of Peres
\cite{PERES}, otherwise the state is separable, i.e.  the negativity
is zero.  It is convenient to use Kimura's method \cite{KIMURA} to
find the $\lambda_{i}$'s.  It is known that there is only one negative
eigenvalue $\lambda_{min}$ of $\rho_{OUT}^{PT}$ in the case of two
dimensional bipartite mixed states (in general dimensions there may
be many).  The eigenvalue $\lambda_{min}$ can be either zero or
negative, depending on the parameters chosen for the unitary
transformation $U$ and on the unknown initial state $\rho_{0, int}$.

Expanding the determinant in the eigenvalue equation we get 
\be
F(\lambda)\equiv det(\rho_{OUT}^{PT}-\lambda)=\lambda^4-a_{1}\lambda^3+a_{2}
\lambda^2-a_{3}\lambda+a_{4}, 
\label{det} 
\ee 
where the coefficients $a_{i}$'s are given by 
\bea
a_{1}&=&1,\nonumber\\
2!a_{2}&=&1-Tr_{path, int}({\rho_{OUT}^{PT}}^2),\nonumber\\
3!a_{3}&=&1-3Tr_{path, int}({\rho_{OUT}^{PT}}^2)+2Tr_{path, int}({\rho_{OUT}^{PT}}^3),\nonumber\\
4!a_{4}&=&1-6Tr_{path, int}({\rho_{OUT}^{PT}}^3)+3[Tr_{path, int}({\rho_{OUT}^{PT}}^2)]^2\nonumber \\
&-&6Tr_{path, int}({\rho_{OUT}^{PT}}^4). 
\eea 
A straightforward computation gives us 
\bea 
a_{1}&=&1,\nonumber\\
2!a_{2}&=&{1-b_{0}^2\over 2},\nonumber\\
3!a_{3}&=&-{3\over 2}b_{0}^2(x^2+y^2),\nonumber\\
4!a_{4}&=&-{3\over 2}b_{0}^2(x^2+y^2)[x^2+y^2+z^2(1-b_{0}^2)].
\eea
It is clear from the fact that $a_{4}>0$ that at least one of the
eigenvalues of the partial transposition of the density operator is
negative unless $x=y=0$, in which case two of them are zero.
Furthermore the roots of ${d^2F(\lambda)\over d \lambda^2}=0$ are
positive because $a_{1}>0$ and $a_{2}>0$.  We see that the two points
of reflection are positive so that only one of the eigenvalues of the
partially transposed density operator is negative as known for a
general two qubit case.

Just for illustration let us consider a special case ($t=0$) for the unitary
transformation, i.e.
\be 
U=\left(\begin{array}{clcr} iz & ix+y \\ ix-y &
-iz\end{array}\right)\;\; ;~~~ x^2+y^2+z^2=1.  
\ee 
The eigenvalues of
$\rho_{OUT}^{PT}$ are then 
\bea
\lambda_{min}&=&-{b_{0}\sqrt{x^2+y^2}\over
2},\nonumber \\
\lambda_{2}&=&{1+b_{0}z\over 2},\nonumber \\
\lambda_{3}&=&{1-b_{0}z\over 2},\nonumber \\
\lambda_{4}&=&{b_{0}\sqrt{x^2+y^2}\over  2},
\eea
and therefore, explicitly using eq.  (\ref{visibility}) for the visibility
$\Gamma=b_0~ z$ and the unitarity condition $x^2+y^2+z^2=1$, the
negativity reads 
\be 
N(\rho)=\sqrt{b_0^2-\Gamma^2}. 
\label{curve} 
\ee
For $t=0$, the geometric phase is $\gamma_g={\pi\over 2}$.  For a fixed
(unknown) input parameter $b_{0}$ the curve (\ref{curve}) is a
quadrant of a circle with the radius $b_{0}$.  Therefore the
entanglement is a decreasing function of the visibility of the
geometric phase.  The largest entanglement $N=b_{0}$ and the smallest
visibility $\Gamma=0$ are obtained for $z=0$.  This means a flip of the spin
on the vertical path, while the opposite is realized when we do
nothing but perform the standard Mach-Zehnder interferometry.  In the
generic case the visibility is $b_{0}$ times the cosine of the angle
by which the spin is tilted by the unitary transformation, while the
entanglement is $b_{0}$ times the sine of it.  The complementarity
comes from the unitarity of the operation $U$.  The pure state case
that we considered before is reproduced if one puts $b_{0}=1$.
For non-zero $t$, we numerically computed the negativity and the
visibility and plot them together as a function of  $t$ in Fig.  4, with a
fixed $b_{0}=0.7$.  We can see qualitatively the same phenomenon as
the special case $t=0$ for each $t$; the negativity is a monotonically
decreasing function of the visibility.
%%%%%%%%%%%%%%%%%%%%%%%%%%%%%%%%%%%%%%%%%%%

\begin{figure}
\includegraphics[width=.9\linewidth]{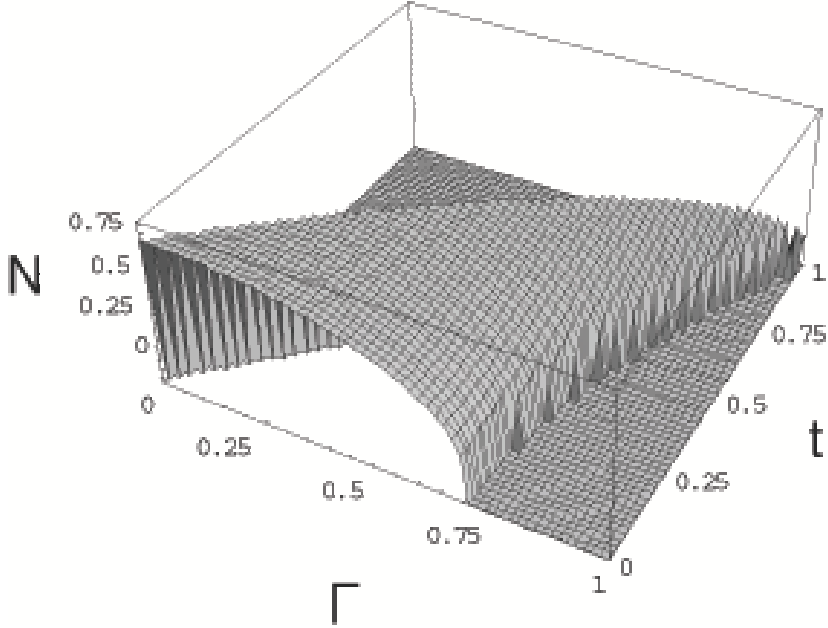}
\caption{
A plot of the negativity N as a function  of the visibility $\Gamma $ and 
the control parameter $t$ of the unitary 
transformation $U$ for a fixed input parameter $b_0 ~(=0.7)$.
The flat area represents the physically accessible states when $U$ is parametrized
by $t>0$ (see eq. (\ref{uparam})). 
For a general mixed state $\rho _{0, int}$ and $t\neq 0$, $N$ is a decreasing 
function of $\Gamma $.}
\end{figure}

%%%%%%%%%%%%%%%%%%%%%%%%%%%%%%%%%%%%%%%%%%%%%%%

Turning back to the eigenvalue equation for $\rho_{OUT}^{PT}$, this can be
more conveniently cast into the form, 
\be
F(\lambda)=\lambda^4-\lambda^3+\alpha
\lambda^2+\beta\lambda-\beta\gamma=0, 
\ee 
where
$\alpha\equiv (1-b_{0}^2)/2$, $\beta\equiv
\{b_{0}^2-[1-(1-b_0^2)\cos^2\gamma_g]\Gamma^2\}/4$ and $\gamma\equiv
(1-\Gamma^2)/4$.  Note that the ranges of the parameters are
$0\leq\alpha\leq 1/2$, $0\leq\beta\leq 1/4$ and $0\leq\gamma\leq 1/4$
and also the fact that the visibility dependence appears only in the
terms $\beta$ and $\gamma$.
Regarding $\lambda$ as the smallest (negative) eigenvalue
of $F(\lambda)$, taking the derivative of $F(\lambda)$ with respect
to $\Gamma$ and using $N(\rho)=-2\lambda$,
we get: 
\be 
{\partial F(\lambda)\over
\partial\lambda}\cdot {dN(\rho)\over d\Gamma} =\Gamma[\beta +(\gamma
+|\lambda|) \delta]>0, 
\ee 
where $\delta\equiv
[1-(1-b_0^2)\cos^2\gamma_g]>0$.  Therefore, since $\lambda$ is the
smallest root of the quartic form, so that the derivative ${\partial
F(\lambda)\over \partial\lambda}$ is negative, we conclude that
${dN(\rho)\over d\Gamma}<0$.
%%%%%%%%%%%%%%%%%%%%%%%%%%%
\subsection{General Mixed State of Two Qubits via CP Maps}
%%%%%%%%%%%%%%%%%%%%%%%%%%%
The model of the previous sections can be generalized by further
applying on both arms of the interferometer (after the two mirrors)
two CP maps (Fig. 5).  The first is given by $W_1$, which acts as
\be
W_1 : \rho_1~\rightarrow ~ \rho_2\equiv (1-p)\rho_1+p A_1\rho_1 A_1^{\dagger},
\label{w1}
\ee 
where $\rho_1\equiv V_0\rho_{IN}V_0^{\dagger}$, with $V_0\equiv
V_MV_CV$, is the density matrix after the mirrors, the parameter
$p\in [0, 1]$ and $A\equiv 1\otimes \sigma_x$.  In other words, $W_1$
may represent the chance of a bit-flip noise in the internal degrees of
freedom with probability $p$.  The second CP map, acting
immediately after $W_1$,
is then given by
\be 
W_2 : \rho_2~\rightarrow ~ \rho_3\equiv
(1-q)\rho_2+q A_2\rho_2 A_2^{\dagger},
\label{w2}
\ee
where now the operator $A_2\equiv \sigma_z\otimes 1$ represents the
chance of a phase noise in the beam (caused, e.g., by an imprecise
notion of the location of the mirrors) with probability $q\in [0,1]$.
It is important to note that, although these two CP maps of course do not
represent the most general situation of possible sources of noise in
the system (there might be other kinds of noise, and in different
locations in the interferometer), the model we consider is already
sufficient to realize a two qubit mixed state with nine parameters
($\vec{b}_0, \chi, p, q$ and $t, x, y, z$ modulo the unitarity
constraint for $U$), i.e.  the most arbitrary four dimensional
density matrix once
the local $SU(2)\times SU(2)$ operations are modded out.

The mixed state finally goes through the last beam splitter,
$\rho_{OUT}\equiv V_H\rho_3V_H^{\dagger}$ and one thus obtains the
output parametrized as 
\be 
\rho_{OUT}\equiv {1\over 4}[I\otimes I +
\vec{a}\cdot \vec{\sigma}\otimes I + I\otimes \vec{b}\cdot
\vec{\sigma} + c_{i j}\sigma_i\otimes \sigma_j],
\label{rhoout}
\ee
where
\bea
\vec{a}&\equiv& \alpha_q[-(t\sin \chi -b_0 z\cos \chi)\vec{e}_y +
(t\cos \chi +b_0 z\sin \chi)\vec{e}_z] ,\nonumber \\
\vec{b}&\equiv & b_0\{(-ty+xz)\vec{e}_x+\alpha_p[(tx+yz)\vec{e}_y
+(t^2+z^2)\vec{e}_z]\},\nonumber \\
\vec{c}_x&\equiv & b_0\{(-ty+xz)\vec{e}_x+\alpha_p[(tx+yz)\vec{e}_y
-(x^2+y^2)\vec{e}_z]\},\nonumber \\
\vec{c}_y&\equiv& \alpha_q\{(x\cos \chi +b_0 y\sin \chi)\vec{e}_x +
\alpha_p[(y\cos \chi -b_0 x\sin \chi)\vec{e}_y
\nonumber \\
&+&(z\cos \chi -b_0 t\sin \chi)\vec{e}_z]\}] ,\nonumber \\
\vec{c}_z&\equiv& \alpha_q\{(x\sin \chi -b_0 y\cos \chi)\vec{e}_x +
\alpha_p[(y\sin \chi +b_0 x\cos \chi)\vec{e}_y
\nonumber \\
&+&(z\sin \chi +b_0 t\cos \chi)\vec{e}_z]\}].
\label{parameters}
\eea
We have here chosen, without loss of generality, $\vec{b}_0=b_0\vec{e}_z$,
and we have redefined for simplicity the parameters $(p, q)$ in terms
of $\alpha_p\equiv 1-2p$ and $\alpha_q\equiv 1-2q$ ($\alpha_{p, q}\in
[-1, 1]$).  Moreover, we have decomposed the matrix $c_{ij}$ appearing
in the last term of eq.  (\ref{rhoout}) as the column vector $c\equiv
(\vec{c}_x, \vec{c}_y,\vec{c}_z)^T$.

%%%%%%%%%%%%%%%%%%%%%%%%%%%%%%%%%%%%
\begin{figure}
\includegraphics{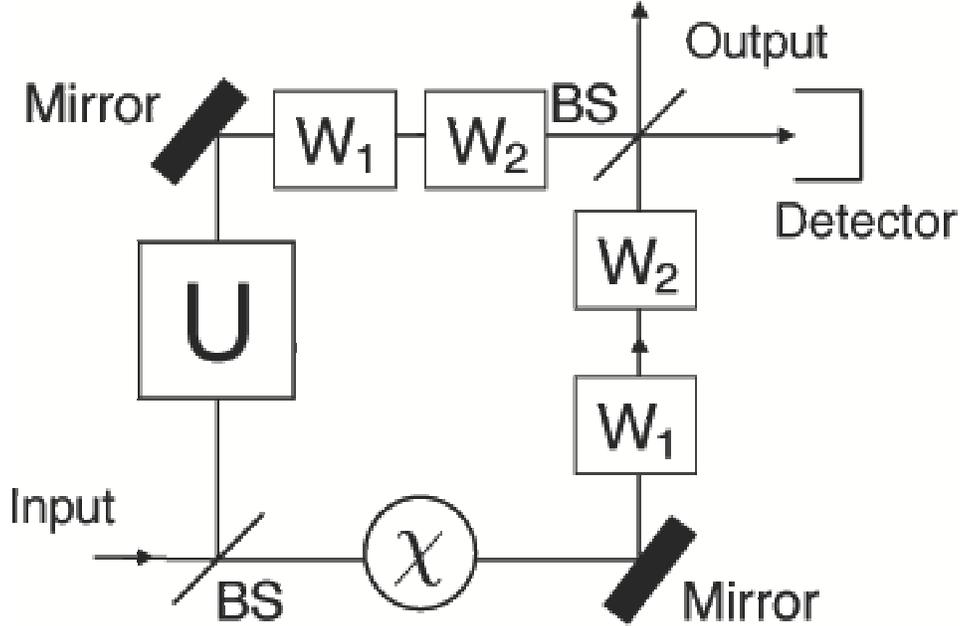}
\caption
{
Sj\"oqvist's interferometer model with the extra CP maps, $W_1$ and $W_2$.
With probability $p$ the spin state of the particles is bit-flipped ($W_1$) 
and with probability $q$ the beam state gets a phase shift $(W_2)$. 
}
\end{figure}
%%%%%%%%%%%%%%%%%%%%%%%%%%%%%%%%%%%%

The probability to detect the particle in the horizontal direction
after the last beam splitter is still given by eq.  (\ref{int1}), but
now the visibility is scaled down by the factor $\alpha_q$, i.e.  
\be
\Gamma=|\alpha_q|\sqrt{t^2+b_0^2z^2},
\ee 
and the geometric phase
(\ref{geomphase}) also acquires an extra contribution from $arg
(\alpha_q)$.  We can now proceed to evaluate the negativity as our
measure of entanglement.  As an example, we consider the analytically
solvable model where the parameters $t$ and $x$ are chosen to be zero.
In this case, the coefficients $a_i$'s in the eigenvalue equation
(\ref{det}) for the partial transpose of the output density matrix
can be compactly written as

\bea
a_2&=&{1\over 4}[(1-b_0^2)+2(1+b_0^2)(\mu_p+\mu_q)-8\mu_p\mu_q],
\nonumber \\
a_3&=&{1\over
4}\{-b_0^2(1-z^2)+[1+b_0^2(3-4z^2)](\mu_p+\mu_q)
\nonumber \\
&-&4[1+2b_0^2(1-2z^2)]\mu_p\mu_q\},
\nonumber \\
a_4&=&{1\over
16}\{-b_0^2(1-z^2)(1-b_0^2z^2)[1-4(\mu_p+\mu_q)]+(1-b_0^2)^2(\mu_p^2+_mu_q^2)
\nonumber \\
&+&2[1-8b_0^2(1-z^2)-b_0^4(1-8z^2+8z^4)]\mu_p\mu_q
\nonumber \\
&-&8(1-b_0^2)\mu_p\mu_q(\mu_p+\mu_q)+16\mu_p^2\mu_q^2\},
\label{kimura}
\eea
where the effects of the non zero probabilities for the bit-flip and
phase noise are explicitly shown through the parameters $\mu_p\equiv
(1-\alpha_p^2)/4=p(1-p)$ and $\mu_q\equiv (1-\alpha_q^2)/4=q(1-q)$,
and the dependence on the visibility must be seen through the substitution
$z^2=\Gamma^2/\alpha_q^2b_0^2$.
It is then immediate to calculate the eigenvalues of the partial transpose
$\rho_{OUT}^{PT}$ as
\bea
\lambda_1&=&{1\over 4}[1-\alpha_p\alpha_q
-\sqrt{b_0^2(\alpha_p+\alpha_q)^2-4\Gamma^2\alpha_p/\alpha_q}],
\nonumber \\
\lambda_2&=&{1\over 4}[1+\alpha_p\alpha_q
-\sqrt{b_0^2(\alpha_p-\alpha_q)^2+4\Gamma^2\alpha_p/\alpha_q}],
\nonumber \\
\lambda_3&=&{1\over 4}[1-\alpha_p\alpha_q
+\sqrt{b_0^2(\alpha_p+\alpha_q)^2-4\Gamma^2\alpha_p/\alpha_q}],
\nonumber \\
\lambda_4&=&{1\over 4}[1+\alpha_p\alpha_q
+\sqrt{b_0^2(\alpha_p-\alpha_q)^2+4\Gamma^2\alpha_p/\alpha_q}].
\label{eigenvalues}
\eea
After some algebra (which we omit here for the sake of simplicity),
one can show that for a certain region of the parameters $(\alpha_p,
\alpha_q)$ (i.e., $(p, q)$) such that $\alpha_p\alpha_q>0$, the
minimal and negative eigenvalue is given by $\lambda_1$, while for
another region for which $\alpha_p\alpha_q<0$ the minimal and negative
eigenvalue is $\lambda_2$.  As an example, in the case in which there
is only the chance of bit-flip noise (i.e., $q=0$), we get that
$\lambda_1<0$ for $0<p<p_-<1/2$ and $\lambda_2<0$ for $1/2<p<p_+<1$,
where $p_-\equiv
\sqrt{b_0^2-\Gamma^2}(\sqrt{1-\Gamma^2}-\sqrt{b_0^2-\Gamma^2})/(1-b_0^2)$
and $p_+\equiv
\sqrt{1-\Gamma^2}(\sqrt{1-\Gamma^2}-\sqrt{b_0^2-\Gamma^2})/(1-b_0^2)$.

For general $(p, q)$, unifying the cases with different signs for the product
$\alpha_p\alpha_q$, we then obtain the negativity as
\be
N={1\over 2}[-1+|\alpha_p||\alpha_q|+\sqrt{b_0^2(|\alpha_p|+|\alpha_q|)^2
-4\Gamma^2|\alpha_p|/|\alpha_q|}].
\label{negativity}
\ee
As a consequence, similarly to the case studied in the previous paragraphs,
we obtain the following two results.
First, in the regions where the negativity is non zero, i.e.  where the
output mixed state is entangled, there is a simple geometric
relationship between the visibility and the negativity, which can be
seen by rewriting eq.  (\ref{negativity}) as
\be
{N'^2\over a^2}+{\Gamma^2\over b^2}=1,
\label{conic}
\ee
which clearly represents an ellipse centered at $N'\equiv
N+(1-|\alpha_p||\alpha_q|)/2=\Gamma=0$ with semiaxis $a\equiv
b_0(|\alpha_p|+|\alpha_q|)/2$ and $b\equiv \sqrt{|\alpha_q|/|\alpha_p|}a$.
In other words, the negativity is maximum, $N=[b_0(|\alpha_p|+|\alpha_q|)-
(1-|\alpha_p||\alpha_q|)]/2$ when the the visibility is zero, and
vice versa when the negativity is minimum, i.e.  zero, the visibility
is maximum, ranging as $\Gamma\in [\{b_0^2(|\alpha_p|+|\alpha_q|)^2-
(1-|\alpha_p||\alpha_q|)^2|\alpha_q|/|\alpha_p|\}^{1/2}/2, |\alpha_q|b_0]$.
In Fig.  6 we numerically plotted the values of the negativity $N$ as a
function of the visibility $\Gamma$ and the control parameter $t$
for the case of a bit-flip noise model with fixed
$x=0.4$ and the set of probabilities $p=0.4$
and $p=0.5$.  The second, related, result is again the complementarity
between the visibility and the negativity, which is mathematically
expressed as $d N/d \Gamma
=-|\alpha_p|/[4|\alpha_q|\Gamma\sqrt{b_0^2(|\alpha_p|+|\alpha_q|)^2
-4\Gamma^2|\alpha_p|/|\alpha_q|}<0$.  As the careful reader will have
already noticed, our extended interferometer model with CP maps has
also another property, i.e.  the mixed output states can be either
entangled or separable, depending on the values chosen for the control
parameters $p, q$, $t, x, y, z$ and $\chi$.  This had to be expected
since our model, equipped with the full set of nine parameters, is able
to represent the most general state of a two qubits mixed system.  In
particular, the boundary between the region with entangled states and
that with separable ones is fixed by the condition $a_4=0$.  In the
simple model in which there is only a bit flip noise, i.e.  $q=0$, it
is immediate to check that the entangled states are given for
$0<p<p_-$ or $p_+<p<1$ (obviously, these are the same parameter ranges
for which one of the eigenvalues of $\rho_{OUT}^{PT}$ is negative,
i.e.  for which the negativity is non zero), while the separable
states are for $p_-<p<p_+$.  In other words, at least within our model,
the more random the noise in the qubit representing the internal degree of 
freedom, the less entangled the global mixed state appears to be.
%%%%%%%%%%%%%%%%%%%%%%%%%%%%%%%

\begin{figure}
\includegraphics[width=1.0\linewidth]{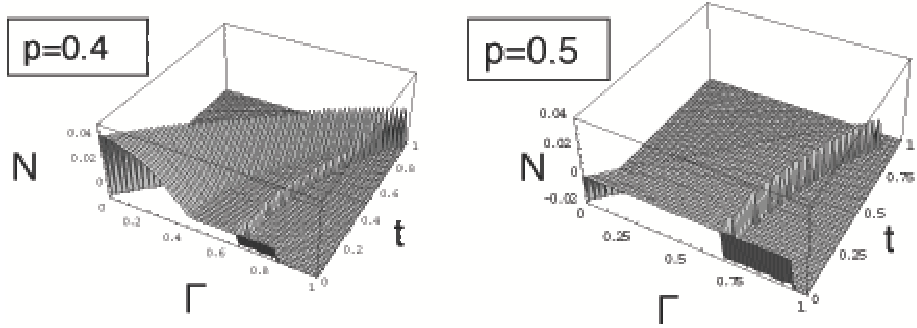}
\caption
{
A plot of the negativity N as a function  of the visibility $\Gamma $ and 
the control parameter $t$ of the unitary 
transformation $U$ for the output of the Sj\"oqvist's interferometer model 
with the extra CP map representing a bit-flip noise, $W_1$ $(q=0)$.
The input parameter $b_0=0.7 $, the unitary transformation control
parameter $x=0.4$ and the CP map parameter $p=0.4 (0.5)$ are fixed.
The lowest flat areas in both graphs represent the regions of the 
physically inaccessible states for the unitary transformation control parameter 
$t>0$ (see eq. (\ref{uparam})).
Note the existence of a boundary between regions with separable states 
$(N=0)$ and entangled states $(N>0)$ 
when $p=0.4$. In this case $N(>0)$ is a decreasing function of $\Gamma $. 
When $p=0.5$ there are only separable states.
}
\end{figure}

%%%%%%%%%%%%%%%%%%%%%%%%%%%
\section{Entanglement of Pure State vs. Visibility of Interference}
%%%%%%%%%%%%%%%%%%%%%%%%%%%%
The entanglement measure of a bipartite pure quantum system is well
established and quantified as the von Neumann entropy of the reduced
density operator given by tracing over the internal states,
\be
\rho_{red}=Tr_{int}(\rho_{OUT})
={1\over 2}[I+\Gamma(\sin(\chi-\phi)\sigma_{y}+\cos(\chi-\phi) \sigma_{z})].
\ee 
This state is depicted in the Bloch ball (Fig.  7) as a vector
with length given by the visibility $\Gamma$ and tilted by the angle
$\chi-\phi$ from the z-axis.  The entanglement measure is the von
Neumann entropy, 
\be
E=S(\rho_{red})=-\lambda_{+}\log\lambda_{+}-\lambda_{-}\log\lambda_{-},
\ee 
where $\lambda_{\pm}$ are the eigenvalues of the reduced density
matrix $\rho_{red}$, expressed in terms of the visibility $\Gamma$ as
\be 
\lambda_{\pm}={1\pm \Gamma \over 2}.  
\ee 
This implies that the
more visible the interference, the less entangled the state (Fig.  8).

%%%%%%%%%%%%%%%%%%%%%%%%%%%%%%%%%%%%%%%%%%%%%%%

\begin{figure}
\begin{center}
\includegraphics[width=.6\linewidth]{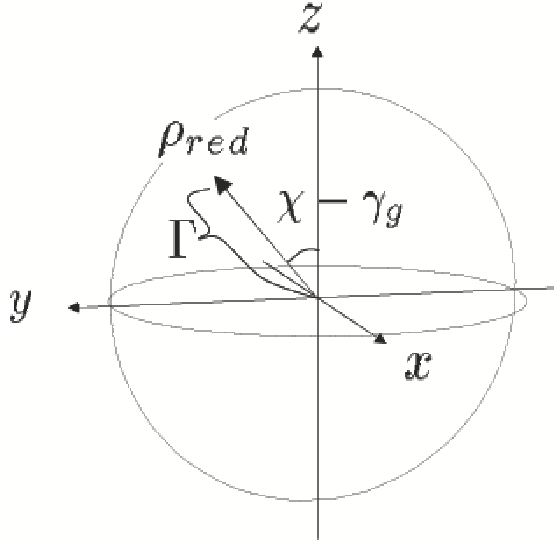}
\end{center}
\caption{
The reduced density matrix as a vector in the Bloch ball. The length of 
the vector is the visibility and 
its polar angle is $\chi -\gamma _g $, with $\gamma _g $ the geometric 
phase. }
\end{figure}
%%%%%%%%%%%%%%%%%%%%%%%%%%%%%%%%%%%%%%%%%

 \begin{figure}
\begin{center}
  \includegraphics{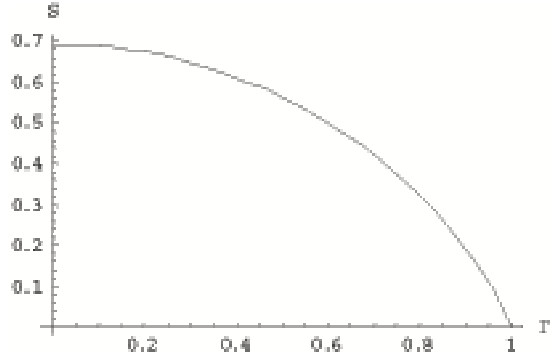}
\end{center}  
\caption{The von Neumann entropy $S$ of the reduced density matrix 
(the entanglement measure for the pure state) is a decreasing function of 
the visibility 
$\Gamma $ of the interference.}
  \end{figure}

%%%%%%%%%%%%%%%%%%%%%%%%%%%%%%%
\section{Complementarity for Pure Bipartite States in Arbitrary Dimensions}

While in the previous section entanglement was achieved by the action of the
interferometer, in this section, for analytical purposes, we want to start with
an already entangled state in Schmidt form and further perform a unitary transformation on the path states.
Of course, in the case of two qubit pure states the two approaches are mathematically equivalent.
In fact, to see this, we can simply choose the initial internal state to be pure, i.e. $\rho_{0, int}
\equiv |\psi_0>_{int}<\psi_0|$, where $|\psi_0>_{int}\equiv a|0>_{int}+b|1>_{int}$, with $a$ and $b$ 
real parameters satisfying $a^2+b^2=1$
(or, equivalently, $\vec{b_0}=(0, 2ab, a^2-b^2)$, with $|\vec{b_0}|=1$) and take $U=\sigma_z$ 
to find that the output state in Eq. 
(\ref{out1}) reads as $\rho_{OUT}=|\psi_{OUT}><\psi_{OUT}|$ with
 \bea 
|\psi_{OUT}>&=&
a(\cos{\chi\over 2}|0>_{path}-i\sin{\chi\over 2}|1>_{path})|0>_{int} \nonumber \\
&+&b(\cos{\chi\over 2}|1>_{path}+i\sin{\chi\over 2}|0>_{path})|1>_{int}.
\eea 
Furthermore, this can be derived from the initial entangled state in Schmidt form
\be
|\psi_{IN}>=a|0>_{path}|0>_{int}+b|1>_{path}|1>_{int} 
\ee 
by operating on the path space with the unitary transformation
$|0>_{path}\rightarrow \cos{\chi\over 2}|0>_{path}-i\sin{\chi\over 2}|1>_{path}$,
$|1>_{path}\rightarrow i\sin{\chi\over 2}|0>_{path}+\cos{\chi\over 2}|1>_{path}$

Now we can proceed in the analysis by observing the first qubit without caring about the second one.
The probability to get the state $|0>_{path}$ is  (cf. Eq. (\ref{int1}))
\be
 P_{|0>_{path}}={1\over 2}[1+(a^2-b^2)\cos\chi],
\ee 
which is a typical interference pattern if we regard the angle $\chi$ as a
control parameter. The visibility of the interference is 
\be 
\Gamma\equiv |a^2-b^2|, 
\ee 
which vanishes when the initial state $|\psi_{IN}>$ is
maximally entangled, i.e.  $a^2=b^2=1/2$, while it becomes maximum when
$|\psi_{IN}>$ is separable, i.e.  $a=0$ or $b=0$.
On the other hand, by choosing the negativity $N=2|ab|$ as our entanglement measure,
we clearly see the complementarity\footnote{A similar expression in terms of the 
distinguishability $D$ of the paths instead of the measure of entanglement $N$, was also discussed and experimentally tested in Ref.  \cite{DURR}.} as
\be 
N^2+\Gamma^2=1.
\ee 
We note that this constraint between the entanglement and the
interference comes from the unitarity $a^2+b^2=1$.
Alternatively, taking the entanglement measure as the von Neumann entropy
we get
\be
E\equiv S(\rho_{red})=-a^2\log a^2-b^2\log b^2, 
\ee
where the
reduced density operator
$\rho_{red}\equiv
Tr_{int}|\psi_{OUT}><\psi_{OUT}|=Tr_{int}|\psi_{IN}><\psi_{IN}|=a^2|0>_{path}<0|+b^2|1>_{path}<
1|$.  The entanglement takes the maximum value $E=1$ when $a^2=b^2=1/2$ and
the minimum value $E=0$ for $a=0$ or $b=0$.  Again, we observe that
the more the state is entangled, the less is the visibility of the
interference and vice versa.\footnote{Actually, it is known that for  bipartite pure states the
von Neumann entropy is the unique measure for entanglement \cite{DHR}.
The negativity does not satisfy the partial additivity condition required in Ref. \cite{DHR}
and therefore does not coincide with the von Neumann entropy.}
   
One can also ask to what extent the complementarity holds for general dimensions.
Consider a bipartite system initially in the state $|\Psi_{IN}>\in {\cal H}_{path}\otimes {\cal H}_{int}$ of 
dimensions $n\times n$ given in a Schmidt decomposition form:
\begin{equation}
|\Psi_{IN}>=\sum_{m=1}^{n}\sqrt{c_m}|m>_{path}|m>_{int},
\end{equation}
with $c_m\geq 0, ~\sum_{m=1}^{n}c_m=1$ and $|m>_{path}\in {\cal H}_{path},
|m>_{int}\in {\cal H}_{int}$.  Suppose then we perform a local unitary
transformation $U_{path}$ to the state $|m>_{path}\in {\cal H}_{path}$,
\begin{equation}
|\Psi_{IN}>\rightarrow |\Psi_{OUT}>=\sum_{m=1}^{n}\sqrt{c_m}U_{path}|m>_{path}|m>_{int},
\end{equation}
and observe the state belonging to ${\cal H}_{path}$.  The probability
to obtain $|0>_{path}$ is
\begin{equation}
P_{|0>_{path}}(U_{int})=\sum_{m=1}^{n}c_{m}|_{path}<0|U_{path}|m>_{path}|^2.
\end{equation}
The dependence of $P_{|0>_{path}}(U_{path})$ as a function of the control parameters 
identifying $U_{path}$ again exhibits an interference
pattern.  To identify the visibility of the interference pattern
$\Gamma$ we subtract the average of $P_{|0>_{path}}(U_{path})$ over $U_{path}$, i.e.  $\int
P_{|0>_{path}}(U_{path})dU_{path}/\int dU_{path}=1/n$, from the maximum of the intensity,
\begin{equation}
{\Gamma\over 2}\equiv Max_{U_{path}}P_{|0>_{path}}(U_{path})-{1\over n}=c_{max}-{1\over n},
\label{VISI}
\end{equation}
where, without loss of generality,
we have chosen $c_{max}$ to be the greatest Schmidt coefficient
(note that for the simple entangled state given by Eq. (1), 
$c_{max}=a^2$ if $a^2\geq b^2$  and $n=2$, and then $\Gamma=2(a^2-1/2)=a^2-b^2$, as it should be).

This expression has to be compared with the entanglement entropy
\begin{equation}
E=-\sum_{m=1}^{n}c_m \log c_m.
\end{equation}
With the normalization condition for  the Schmidt coefficients
$\sum_{m=1}^{n}c_m =1$ in mind we see that,
fixing all the $c$'s except $c_{max}$ and $c_{k}$,
\begin{equation}
{\partial E\over \partial c_{max}}=-\log{c_{max}\over c_{k}}\leq 0.
\end{equation}
We can thus conclude that the entanglement $E$ is a decreasing function
of $\Gamma$ also for the case of bipartite systems in pure states of arbitrary
dimensions.
A keen reader may realize that an analogous phenomenon is ubiquitous  in
quantum information processing: entanglement tends to hinder efficient
interference.

%%%%%%%%%%%%%%%%%%%%%%%%%%%
\section{Summary and Discussion}
%%%%%%%%%%%%%%%%%%%%%%%%%%%%
We have shown the existence of a complementarity relation between the
visibility of interference and bipartite entanglement for a generic two qubit
mixed state in an interferometric system and for bipartite pure states in arbitrary dimensions.
Explicitly, for bipartite pure states in arbitrary dimensions, the von Neumann
entropy, which is the unique measure for entanglement, is a decreasing function of
the visibility of interference.  
In the case of two qubit mixed states the same complementarity relation holds,
with the negativity taking the role of the entanglement measure.
However, there is no unique measure of entanglement for mixed states \cite{HO},
so that it remains to be seen whether our result is still valid for other measures.
The phenomenon described in this paper can be seen as the
analogue of the complementarity of the Stern-Gerlach experiment and
Young's double slit experiment.  The former is normally performed in
the setting in which the spin and the path are maximally entangled so
that there is no interference pattern, while the polarizations are not
entangled to the paths in Young's interference experiment.  
Another example of the complementarity is the dephasing phenomenon,
because the interferometric system becomes entangled with the environment.
The complementarity may also open up the possibility to quantitatively
measure the entanglement by looking at the visibility of the
interference (see, e.g., Ref. \cite{CARTERET}). 

Interestingly enough the issue of entanglement and interference in basic
science has a direct implication in quantum computation, in the sense that the
complementarity might be a useful tool to analyze its efficiency.  Quantum
computation is usually assumed to start with a standard separable state
and then to make a series of unitary transformations to create a
particular entangled state, i.e. 
\be 
\sum_{k}|k>|f(k)>, 
\ee
so as to obtain the many candidates for the solutions to a given problem
characterized by the function $f(k)$ in a quantum parallel way.

In the case of Shor's algorithm, $f(k)$ is a periodic function of $k$.
Suppose we observe the second state to obtain $f(k*)$. The whole state
collapses to a superposition  
\be 
\sum_{k, f(k)=f(k*)}|k>|f(k*)>. 
\ee 
The period
can be read off by doing a local Fourier transformation on the first
state and then looking at the interference pattern.  We can clearly
see that the observation of the second state cuts the entanglement to
make the interference possible.

In general, to solve meaningfully a complex problem one might need sufficient
entanglement while also requiring sufficient interference to get the result
efficiently. 
It would be intriguing if a kind of complementarity similar to that analyzed in our
work were playing a prominent and decisive role in this process.

We have to caution the reader that the entanglement discussed in the present
work is limited to that of bipartite systems. Jozsa and Linden \cite{JL}
discussed the role of entanglement which spreads over an entire set of
qubits in a quantum computation.  If we can quantify entanglement for
macroscopic systems, it would be very interesting to compare that with the
visibility of interference, e.g.  in Shor's algorithm.

\bigskip
%\newpage
\section*{Acknowledgements}
\bigskip
A.H.'s research was partially supported by the Ministry of Education,
Science, Sports and Culture of Japan, under grant n. 09640341. A.H. and
A.C.  are also supported by the COE21 project on `Quantum Computation Geometry'
at Tokyo Institute of Technology.  
%%%%%%%%%%%%%%%%%%%%%%%%%

\end{document}